\documentclass[referee]{raa}            

\usepackage{graphicx,times}             
\usepackage{natbib}
\usepackage{amssymb,amsmath}

\usepackage{multirow}
\usepackage{natbib}
\usepackage{cite}
\usepackage{diagbox}
\bibpunct{(}{)}{;}{a}{}{,}

\usepackage[a4paper=true,pagebackref=true]{hyperref}
\hypersetup{colorlinks = true, linkcolor = green, anchorcolor = red, citecolor = blue, filecolor = red, pagecolor = red, urlcolor = red}

\begin{document}

   \title{Pulsar Candidate Classification with Deep Convolutional Neural Networks
}

   \volnopage{Vol.0 (20xx) No.0, 000--000}      
   \setcounter{page}{1}          

   \author{Yuan-Chao Wang
      \inst{1,2}
   \and Ming-Tao Li
      \inst{1,2}
   \and Zhi-Chen Pan
      \inst{3,4,5}
   \and Jian-Hua Zheng
      \inst{1,2}
   }

   \institute{National Space Science Center, Chinese Academy of Sciences, Beijing 100190,China; {\it limingtao@nssc.ac.cn}\\
        \and
             University of Chinese Academy of Sciences, Beijing 10004,China;\\
        \and
             National Astronomical Observatories, Chinese Academy of Sciences, Beijing 10012,China;\\
        \and
             Center for Astronomical Mega-Science, Chinese Academy of Sciences, Beijing 100012,China;\\
        \and
             CAS Key Laboratory of FAST, NAOC, Chinese Academy of Sciences, Beijing 100012,China;
\vs\no
   {\small Received~~20xx month day; accepted~~20xx~~month day}}

\abstract{ As the performance of dedicated facilities has continually improved, large numbers of  pulsar candidates are being received, which makes selecting valuable pulsar signals from the candidates challenging. In this paper, we describe the design for a deep convolutional neural network (CNN) with 11 layers for classifying pulsar candidates. Compared to artificially designed features, the CNN chooses sub-integrations plot and sub-bands plot in each candidate as inputs without carrying biases. To address the imbalance problem, a data augmentation method based on synthetic minority samples is proposed according to the characteristics of pulsars. The maximum pulses of pulsar candidates were first translated to the same position, and then new samples were generated by adding up multiple subplots of pulsars. The data augmentation method is simple and effective for obtaining varied and representative samples which keep pulsar characteristics. In experiments on the HTRU 1 dataset, it is shown that this model can achieve recall of 0.962 and precision of 0.963.
\keywords{pulsars: general -- methods: statistical -- methods: data analysis}
}

   \authorrunning{Y.-C. Wang et al.  }            
   \titlerunning{Pulsar Candidate Classification}  

   \maketitle

%
%
\section{Introduction}           
\label{sect:intro}

 Searching for pulsar is an important frontier in radio astronomy. Scientists are paying more attention to pulsars because of their broad impact across physics, astronomy, astronautics \citep{Cordes2004,Lorimer1998, Lyne2004, Hobbs2009, Sheikh2006}, etc.
 Many dedicated surveys have been used to search for more pulsar signals, such as the Parkes Multi-Beam Pulsar Survey (PMPS, \citealp{Manchester2001}), High Time Resolution Universe(HTRU, \citealp{Keith2010}) survey, and so on.
 With the advent of large-scale facilities that can conduct such surveys, such as the Five hundred meter Aperture Spherical radio Telescope (FAST, \citealp{Nan2011}), and the Square Kilometre Array (SKA, \citealp{Smits2009}), weaker pulsar signals can be received, although being mixed with more and more noise or radio frequency interferences (RFIs), which makes it difficult to identify valuable suspected pulsar signal from large numbers of pulsar candidates.
 Researchers have applied many successful methods to select
 pulsar candidates, including manual selection \citep{Stokes1986, Johnston1992}, selection with graphical tools \citep{Faulkner2004,Keith2009}, ranking and scoring approaches \citep{Lee2013} and machine learning methods \citep{Eatough2010,Bates2012,Morello2014, Zhu2014, Lyon2016, Devine2016, Guo2017,Bethapudi2018,Tan2018}.

In related works, supervised machine learning methods have become more significant and the major methods in classifying pulsar candidates.

The first published work which attempted to use a machine learning approach to select candidates is \citet{Eatough2010}. They implemented artificial neural networks (ANN) with 12 designed experimental features as input vectors. Then \citet{Bates2012} and \citet{Morello2014} performed some work to improve the performance by optimizing designed features with ANN.
In these methods, the designed features relied on human experience, and may carry unexpected biases against particular types of pulsar candidates \citep{Morello2014, Lyon2016}. To address these problems, \citet{Lyon2016} and \citet{Tan2018} selected fundamental and statistical features which aimed to minimize biases and selection effects \citep{Morello2014} to provide better generalization performance.

In addition to these approaches with artificially designed features, data-driven methods also play an important role in this field.
 \citet{Zhu2014} developed the Pulsar Image-based Classification System (PICS) by using a group of supervised machine learning approaches.
It produces classification based on image patterns. The inputs are four important diagnostic plots of candidates rather than extracted features.
This avoids possible shortcomings in artificially designed features and relying on excessive information.
It has been validated to have a superior ability at recognition in the PALFA survey pipeline and has discovered six new pulsars.
To address class imbalance problems in pulsar candidates, \citet{Guo2017} used a Deep Convolution Generative Adversarial Network (DCGAN, \citealp{Radford2015}) to generate more candidates and automatically extract deep features at the same time. Then they used deep features to classify data, which helps to make the classifier more accurate.

In this paper, we take a step towards improving performance by the data-driven method. We designed a deep convolutional neural network(CNN) with eight convolutional layers, one flatten layer and two fully connected layers. The inputs are the sub-integrations plot and sub-bands plot in each candidate rather than artificially designed features.
To improve class balancing, we designed a simple and useful approach to synthesize more diverse pulsar candidates. New samples were synthesized by adding up multiple subplots of pulsars after maximum pulses of pulsars were shifted to the same position. We tested our model on the HTRU 1 dataset. The results show that our model can provide satisfactory results on both recall and precision.
This paper is organized as follows:
In Section 2, the dataset applied for training is introduced. Section 3 describes the data augmentation method for pulsar candidates. Section 4 introduces the network architecture and training details. Section 5 presents the experimental results of our model and analyses of its performance. Finally, Section 6 is the conclusions of our work.


\section{Dataset}           
\label{sect:Dataset}

To train and test our model, we need labelled convictive datasets. At present, there are relatively few public labeled datasets.. The most common one is the HTRU 1 dataset
\footnote{http://astronomy.swin.edu.au/$\sim$vmorello/}, produced by \citet{Morello2014}.

This dataset is a part of the outputs from new processing of HTRU intermediate Galactic latitude data \citep{Morello2014}. It contains 1196 pulsars from 521 distinct sources with varying spin periods, duty cycles, and signal to noise ratios(SNRs).
Furthermore, it has 89,995 non-pulsar candidates. It has been examined in some recent works \citep{Morello2014, Lyon2016, Guo2017, Ford2017}. In this paper, we implemented it to train and measure our model.

\begin{figure}[htp]
   \centering
   \includegraphics[width=0.7\textwidth, angle=0]{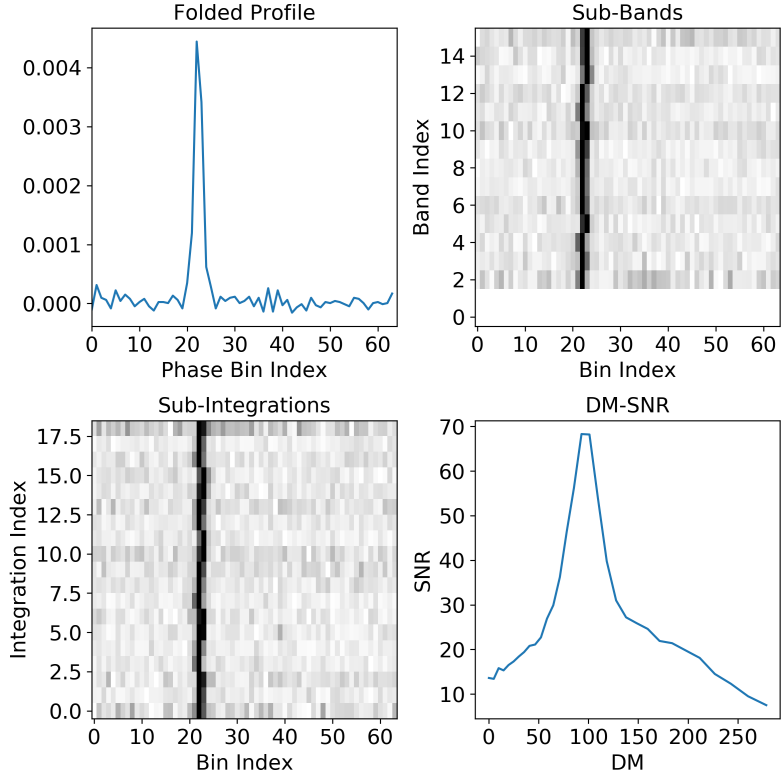}
   \caption{A pulsar example (pulsar\_0023) from the HTRU1 dataset. The four subplots are the folded profile plot (upper left), sub-integrations plot (lower left), sub-bands plot (upper right) and DM-SNR curve (lower right). }
   \label{Figure1}
   \end{figure}

Figure ~\ref{Figure1} is an example of pulsar candidates (pulsar\_0023) in the HTRU 1 dataset with its four most important subplots.
The first one subplot is a folded profile plot. It was obtained by summing the signal in all frequencies and periods. The pulse profile of a typical pulsar would be composed of one or several narrow peaks above the noise floor.
The lower left one is the sub-integrations plot, and it is obtained by summing the data from different frequency channels. It reflects the intensity of the signal during the observation time.
For an ideal pulsar signal, the signal would be observed throughout the observation period, so one or several vertical stripes will form corresponding to the peak positions in the profile curve.
The sub-bands plot is at the upper right. By summing data over all periods, it reflects the intensity of the signal at different frequencies. Since radio pulsars are broadband, there should be one or more vertical stripes in most frequencies.
In the dispersion measure(DM)-SNR curve at the lower right, the SNR as a function of DM is recorded. As the pulse passes through the interstellar medium, it would disperse. The dispersion curve shows the corresponding SNR of the pulse curve when different dispersion values are used for de-dispersion. Therefore, if it is a pulsar signal, the curve will have a peak at the non-zero position, which means the correct value is used for de-dispersion.

In the training process, the inputs are the sub-integrations subplot and sub-bands subplot of each candidates. In HTRU1, the data size of subplots in each candidate may be different. For convenience when extracting features, each subplot data point was resized to the same size, 64*64.
The HTRU1 dataset is very imbalanced (approximately 1:75). As is well known, class imbalance has a negative effect on the performance of the resulting classifiers \citep{He2009, Buda2018}. So, before training our model, it is necessary to address this effect.

\section{Data Augmentation}
\label{sect:Augmen}

In this section, the data augmentation methods that are employed to address class imbalance are introduced. In deep learning, data augmentation techniques, such as rotation, rescaling, shifting, shearing, local warping, adding noise, etc., are frequently applied to increase the diversity of the classes and reduce overfitting of the model.
In this situation, we need more pulsar candidates for training, but the traditional techniques (such as rotation or shearing) are not suitable for pulsar characteristics. Considering this, a simple and specific image synthesis approach is designed to ¡°generate¡± more pulsar candidates.

Our method is an oversampling method based on synthetic minority samples. The details are as follows:
	
(1) Translating the maximum pulse to the same middle position for all pulsar candidates.

(2) Randomly Selecting  three candidates $A$, $B$, $C$ from training pulsar sub-dataset every time.

(3) Adding up their corresponding sub-integrations plots or sub-bands plots with random coefficients.

\begin{equation}
    sub\_int(new)=\alpha \cdot sub\_int(A)+ \beta \cdot sub\_int(B)+\gamma \cdot sub\_int(C)
\end{equation}
\begin{equation}
    sub\_band(new)=\alpha \cdot sub\_band(A)+ \beta \cdot sub\_band(B)+\gamma \cdot sub\_band(C)
\end{equation}

where $\alpha+\beta+\gamma=1$, and $\alpha$ , $\beta$ and $\gamma$ are randomly selected.

(4) Repeat steps 1-3 until enough samples are produced.

\begin{figure}
   \centering
   \includegraphics[width=\textwidth, angle=0]{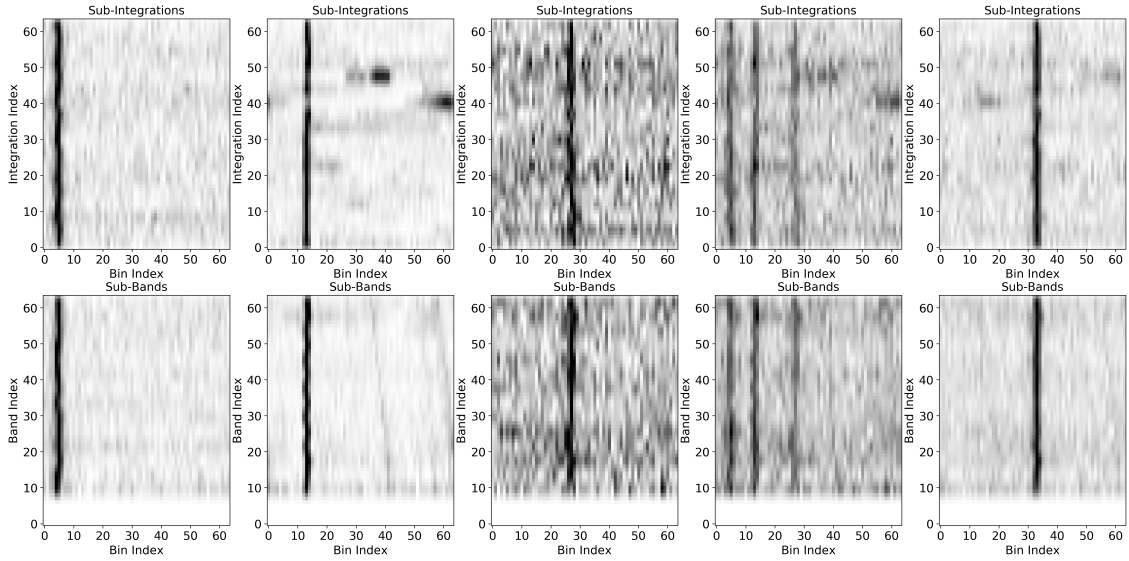}
   \caption{An example of image synthesis. From left to right, are pulsar\_25, pulsar\_62, pulsar\_147, one of the synthetic samples without firstly adjusting pulse positions, and one of the synthetic samples after firstly adjusting pulse positions ($\alpha=\beta=\gamma=\frac{1}{3}$). The upper row displays the sub-integrations plots and the lower row depicts the sub-bands plots. }
   \label{Figure2}
   \end{figure}

In this way, some new pulsar candidates were generated as supplementary positive samples in training.
It should be noticed that making an adjustment to pulse positions before selecting samples in step 1 is necessary to keep the characteristics of pulsars.
Because positions of pulse phase for different pulsars varies, adding up multiple subplots in the above method may lead to multi-pulse plots, which are mussy and may change the sample distribution (an example is shown in Figure ~\ref{Figure2} column 4).
The inputs are subplots with the same size, so this method did not use features of varied pulse width directly.
In each subplot, vertical stripe width is related to its pulse width.
When adding up different subplots, this aspect will influence the intensity or width of vertical stripes, and enhance diversity of training data in a way.
In addition, to further increase diversity, pulse position of positive pulsar candidates can be shifted randomly, when the generative process has finished.

Figure ~\ref{Figure2} shows a synthetic sample. In this figure, the upper row is sub-integrations plots and the lower row is the corresponding sub-bands plots. The first three columns are data from three pulsars. The fourth column is one of their failed synthetic samples without making adjustment to pulse positions before applying steps 2 and 3. In the last column, one of the correct synthetic samples is displayed. The new synthetic sample exhibits different noises and signal strengths that correspond with the original samples.

\begin{figure}[htp]
   \centering
   \includegraphics[width=0.8\textwidth, angle=0]{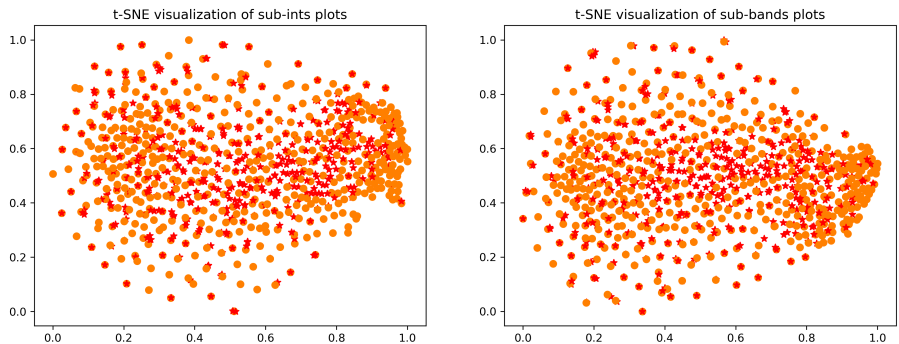}
   \caption{{Visualization of plots using t-SNE. `$\bullet$' marks the 500 original pulsar data, while `$\star$' signifies 500 synthetic samples. The left panel is based on sub-integrations, and the right one is based on sub-bands. We can find that these synthetic samples vary slightly from the original data.}}
   \label{Figure3}
   \end{figure}

In order to investigate the distribution of new samples, Figure ~\ref{Figure3} depicts a visualization of 500 original pulsar samples and 500 synthetic samples shown by using t-distributed Stochastic Neighbor Embedding (t-SNE, \citealp{Maaten2008}). The t-SNE is a non-linear dimensionality reduction algorithm utilized for visualization. In Figure 3, the corresponding high-dimensional data are reduced to two dimensions for visualization. These synthetic samples keep the characteristics of original samples while having some changes in strength of signals or noises.

\section{Model}
\label{sect:Model}

\subsection{Convolutional Neural Networks}

In this section, CNNs are briefly introduced. CNNs have achieved outstanding performance in computer vision, speech recognition and natural language processing in recent years \citep{Michael2015}.
Many CNN modules have been designed for different tasks, such as AlexNet \citep{Krizhevsky2012}, VGG \citep{Simonyan2015}, ResNets \citep{He2016}, CapsuleNet \citep{Sabour2017}, etc.
These models have become deeper and more complex to meet the needs of different tasks.

A typical CNN has convolution layers, pooling layers and fully connected layers.
Convolution layers are used to extract features. In convolution layers, convolution is a specialized kind of linear operation \citep{Goodfellow2016} on feature maps from previous layers with learnable kernels.
Then an activation function (such as sigmoid, hyperbolic tangent, softmax, rectified linear unit or Leaky ReLU) makes a linear or non-linear operation on the output of the kernels to form the output feature maps.
In this way, each of the output feature maps can be combined with more than one input feature map \citep{Alom2018}.
The new feature maps will be the inputs of the next layer. The convolution can be expressed as
\begin{equation}
    x^l_j=f(\sum_{i\in M_j} w^l_{ij}*x^{l-1}_i+b_j^l)
\end{equation}
where $x_j^l$ is the output of the current layer, $x_j^{l-1}$ is the previous layer output, $w_{ij}^l$ is the kernel weight for the present layer, $b_j^l$ is the biases for the current layer, and $f$  represents the activation function.

Pooling layers are usually sandwiched between convolution layers. They represent a down sampled operation (such as average pooling or max-pooling) on the input maps to compress data and reduce overfitting.
Fully connected layers are usually at the ends of CNNs. They are used to integrate extracted features from the preceding layers and output the score of each class. In general, there would be a flatten layer before fully connected layers are used for flattening any input tensor into a vector.

\subsection{Model architecture}
\begin{figure}[htp]
   \centering
   \includegraphics[width=\textwidth, angle=0]{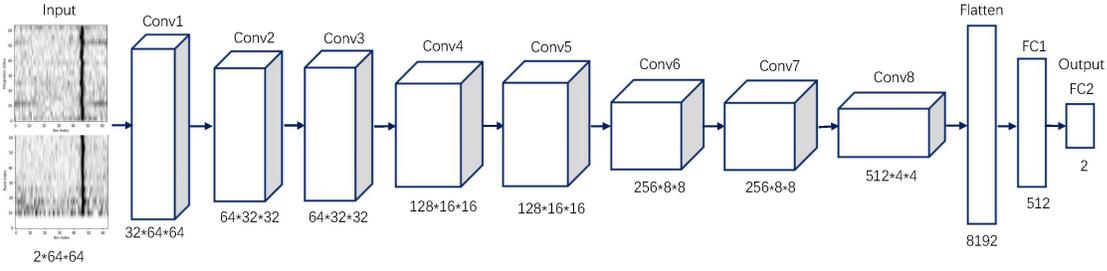}
   \caption{The CNN architecture implemented in this design. This model takes in grayscale subplots with size $64*64*2$ and outputs the classification probability scores for two classes. The network has a total of 11 layers with eight convolutional layers, one flatten layer and two fully connected layers. }
   \label{Figure4}
   \end{figure}

For this classification task, a deep CNN model was designed with eight convolutional layers, one flatten layer and two fully connected layers. The network takes sub-integrations plots and sub-bands plots as inputs. The input size is $64*64*2$, and outputs a tensor of $1*2$, which means the predicted probability of a non-pulsar or pulsar, respectively. The network architecture is shown in Figure ~\ref{Figure4}.

In the convolutional layers, Leaky ReLU was chosen as the activation function, and ReLU and sigmoid activation was used for the two fully connected layers respectively.

In order to reduce overfitting, some useful techniques were used, including batch normalization \citep{Ioffe2015}, dropout \citep{Hinton2012} and L2 regularization.
We added batch normalization with a momentum of 0.9 before each activation layer. This has been demonstrated to help accelerate training processes \citep{Alom2018}.
Then, dropout was applied with alpha of 0.25 after each activation in all convolutional layers and an alpha of 0.55 for the first fully connected layer.
The key idea of dropout is to randomly delete some units during training \citep{Srivastava2014} with a fixed probability $($alpha$)$.
In addition, L2 regularization incorporates a regularization term on the cost function for weight decay. In this model, we used L2 regularization with a parameter of 0.02.

\subsection{Details}
The loss function utilized for optimizing the model is a binary cross-entropy cost function. The binary cross-entropy cost function can be expressed as:
\begin{equation}
    L=-\frac{1}{n} \sum_{x}\left[y \ln \hat{y}+(1-y) \ln (1-\hat{y})\right]
\end{equation}

where the sum is over all training inputs $x$, $n$ is the total number of training data, and $y$ is the actual label with $\hat{y}$ being the corresponding predicted label.

Our model is optimized using a first-order gradient-based optimization algorithm called Adam with default parameters \citep{Kingma2015} and a batch size of 512. In each epoch, 512 candidates are randomly selected as a batch from the training set to train the model and update parameters, which will repeat until all training data have been used. The implementation is based on Python 3.6 and Keras 2.2.4.

 \section{Results}
\label{sect:Results}

\subsection{Evaluation Metric}	
 To provide comprehensive assessments of model performance on this imbalanced dataset, we examined three metrics: recall, precision and F1. These metrics are defined as follows:
 \begin{equation}
Recall=\frac{TP}{TP+FN}\\
\end{equation}
\begin{equation}
Precision=\frac{TP}{TP+FP}\\
\end{equation}
\begin{equation}
F1=2 \cdot \frac{Recall\cdot Precision}{Recall+Precision}
\end{equation}

where TP is True Positive, TN as True Negative. These mean positive data (pulsar candidates) or negative data (RFI) being correctly recognized, respectively.
FN is False Negative while FP is False Positive, and they represent the part of the positive data or negative data being incorrectly labeled, respectively.
Recall indicates the fraction of pulsars correctly being recognized, and precision gives a measure of the fraction of pulsars in predicted positive data. F1 is a harmonic mean of recall and precision.
To identify the largest possible fraction of pulsars while returning a minimal amount of mislabeled noise or RFI \citep{Morello2014}, a good model should provide a high recall as well as a high precision.

\subsection{Experiment on HTRU1}
Before training, the dataset was divided randomly into three parts: training data (60\%), validation data (20\%) and testing data (20\%).
The first and second part of the data were used to train the model while the testing part was used to test the performance of the trained model.
The proposed data augmentation in section 3 only acted on the training data.
With these augmentation methods, the number of pulsar sub-datasets for training was extended by a factor of 30. So, in each training process, the training set had 75507 samples and the validation set employed for hyperparameter selection had 18240 samples. More details are provided in Table~\ref{Tab:1}.

\begin{table}
\begin{center}
\caption[]{ The number of samples in three parts.}\label{Tab:1}


 \begin{tabular}{clcl}
  \hline\noalign{\smallskip}
Part &  All    & Pulsar & Non-pulsar                    \\
  \hline\noalign{\smallskip}
Training set	&75507	&21510	&53997 \\ 
Validation set	&18240	&240	&18000 \\
Testing set	&18240	&239	&17799          \\
  \noalign{\smallskip}\hline
\end{tabular}
\end{center}
\end{table}

\begin{figure}[htp]
   \centering
   \includegraphics[width=0.7\textwidth, angle=0]{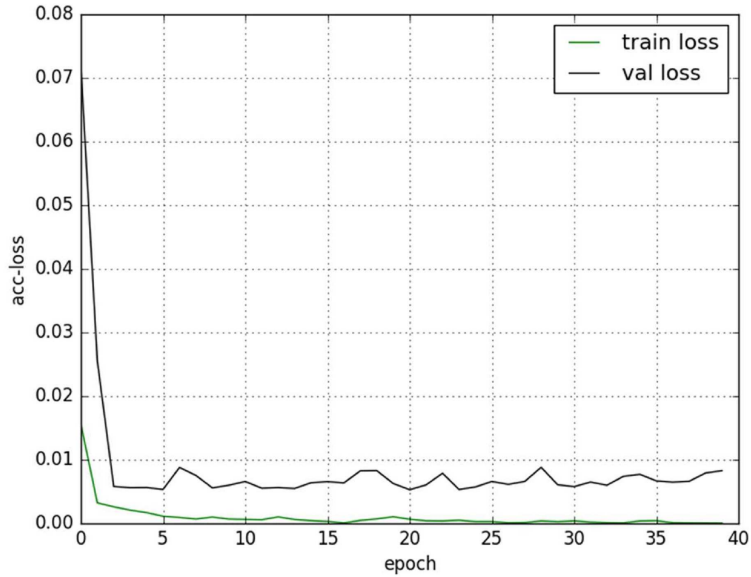}
   \caption{The learning curve of one training process. }
   \label{Figure5}
   \end{figure}

The training operation was processed for 40 epochs, and the learning curve of one training process is shown in Figure ~\ref{Figure5}.
The entire training process was repeated 30 times, with different random data partitions, to obtain an average score for candidates.

Our results are listed in Table~\ref{Tab:2}. On testing data, our CNN model without synthetic samples (only copying minority pulsar candidates 32 times when training), can achieve recall of 0.851 and precision of 0.848(shown in Table~\ref{Tab:2} as DCNN-C). When the model was trained with data augmentation, our model showed excellent recall, precision and F1.
It can achieve recall of 0.962 while precision is 0.963 (shown in Table~\ref{Tab:2} as DCNN-S).
We think that the improvement is based on diversity of the synthetic samples.
We also list some representative results of the data-driven method from \citet{Guo2017} as a contrast.
In \citet{Guo2017}, they tested the CNN model with the same architecture as in \citet{Zhu2014}.
Meanwhile, they introduced the DCGAN+SVM method, in which they used DCGAN to generate synthetic samples and extract deep features and a support vector machine (SVM) to produce better classifications with deep features.
With the help of these synthetic samples, our CNN model has improved approximately 0.6\% in recall and 1.2\% in precision compared to the CNN in \citet{Guo2017}.
It can reach the level of using the DCGAN model.
Though a Generative Adversarial Network(GAN) can generate new samples by adversarial training, it is difficult to train.
In comparison, our synthetic method is simpler than using GAN to generate samples.

\begin{table}
\begin{center}
	\centering
	\caption{ Performance of different methods on HTRU 1 dataset.}\label{Tab:2}
	
\begin{tabular}{ccccc}
\hline
  \textbf{Reference}&  \textbf{Method}	& \textbf{Recall}	& \textbf{Precision}	& \textbf{F-Score} \\
\hline

 \multirow{4}*{\citet{Guo2017}} &CNN-1& 0.956	& 0.950	& 0.953\\
                                & CNN-2	&0.953	&0.951	&0.952  \\
                                &DCGAN-SVM-1	&0.963	&0.965	&0.964 \\
                                &DCGAN-SVM-2	&0.966	&0.961	&0.963\\
\hline
\multirow{2}*{Our method}	&DCNN-C	&0.851	&0.848	&0.849\\
	                        &DCNN-S	&0.962	&0.963	&0.962\\
\hline
\end{tabular}
\end{center}
\end{table}

\subsection{Analyses}	

\begin{figure}[htp]
   \centering
   \includegraphics[width=0.8\textwidth, angle=0]{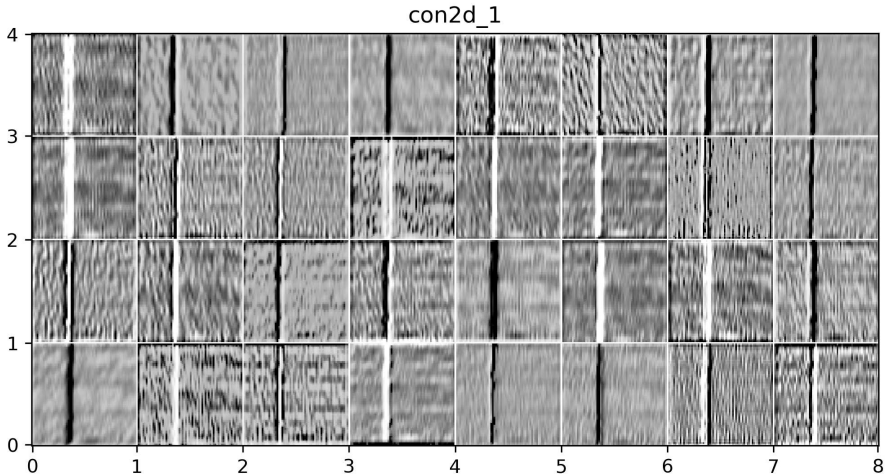}
   \caption{Visualization of 32 feature maps from the first convolutional layer of the network. Each row has eight feature maps. }
   \label{Figure6}
   \end{figure}

\begin{figure}[htp]
   \centering
   \includegraphics[width=0.8\textwidth, angle=0]{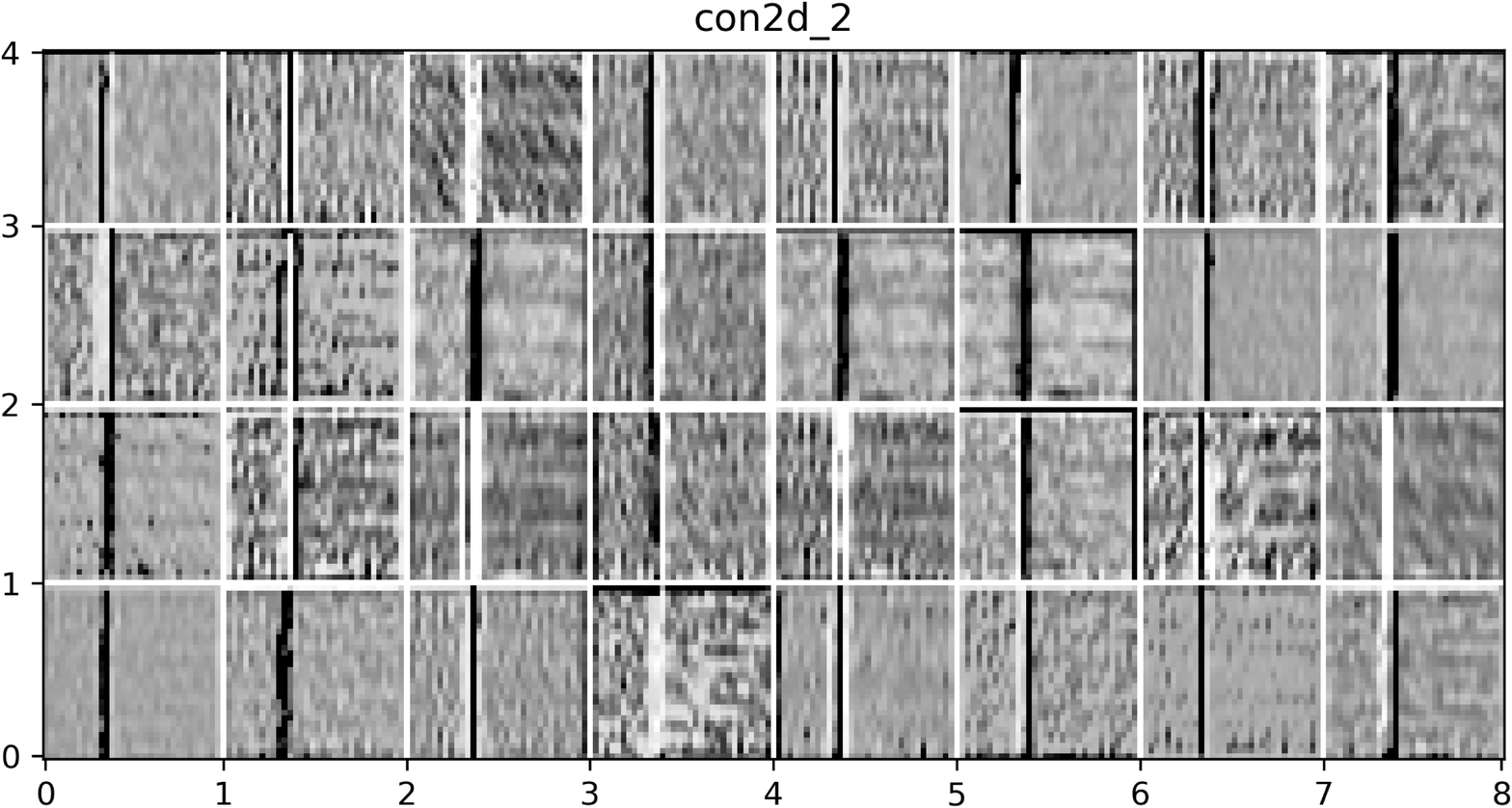}
   \caption{Visualization of 32 feature maps from the second convolutional layer of the network.Each row has eight feature maps.
  }
   \label{Figure7}
   \end{figure}

\begin{figure}[htp]
   \centering
   \includegraphics[width=0.8\textwidth, angle=0]{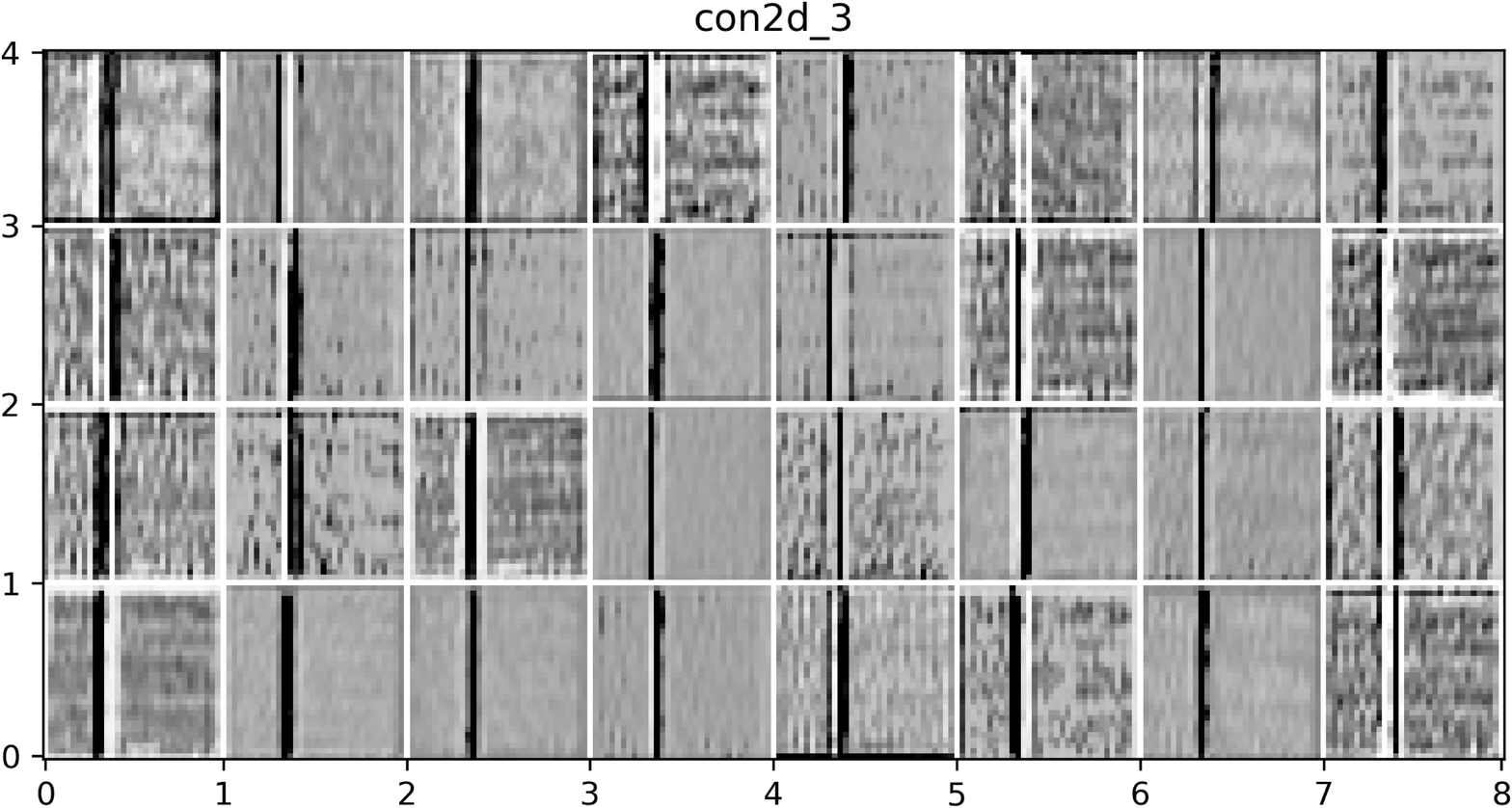}
   \caption{Visualization of 32 feature maps from the third convolutional layer of the network. Each row has eight feature maps.}
   \label{Figure8}
   \end{figure}

\begin{figure}[htp]
   \centering
   \includegraphics[width=0.8\textwidth, angle=0]{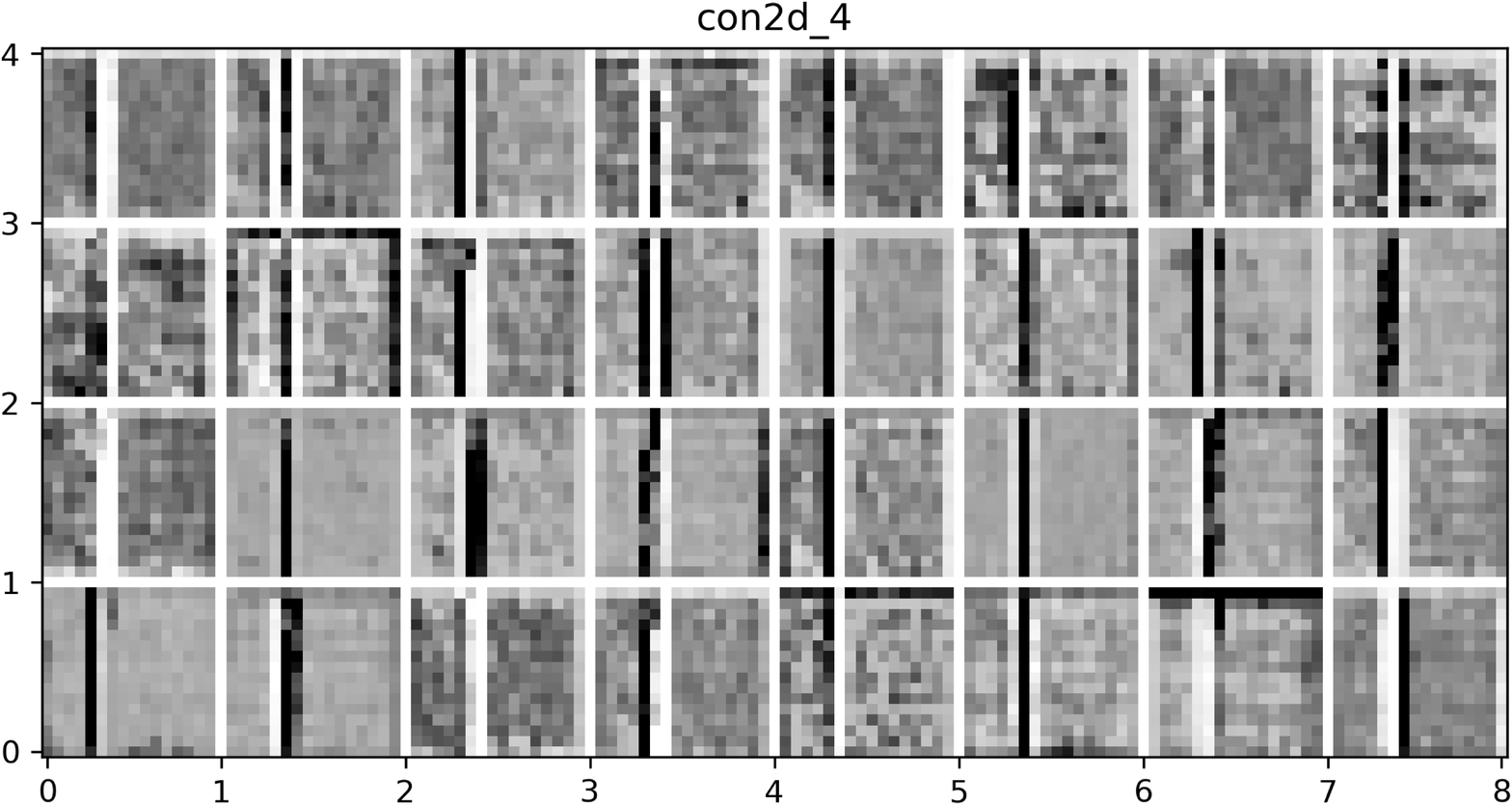}
   \caption{Visualization of 32 feature maps from the fourth convolutional layer of the network. Each row has eight feature maps.}
   \label{Figure9}
   \end{figure}

To analyze features extracted by this model,
intermediate convolution layer outputs of pulsar\_0023 were visualized.
Figure ~\ref{Figure6} to Figure ~\ref{Figure9} depict some feature maps of convolution layers 1 to 4.
By comparing them, we can find that it extracts detailed features in shallow layers, and as the layers deepen, feature maps become more abstract and highlight the most important vertical stripe areas.

\begin{figure}[ht]
  \centering

  \begin{minipage}[t]{0.8\linewidth}
  \centering
   \includegraphics[width=0.8\textwidth]{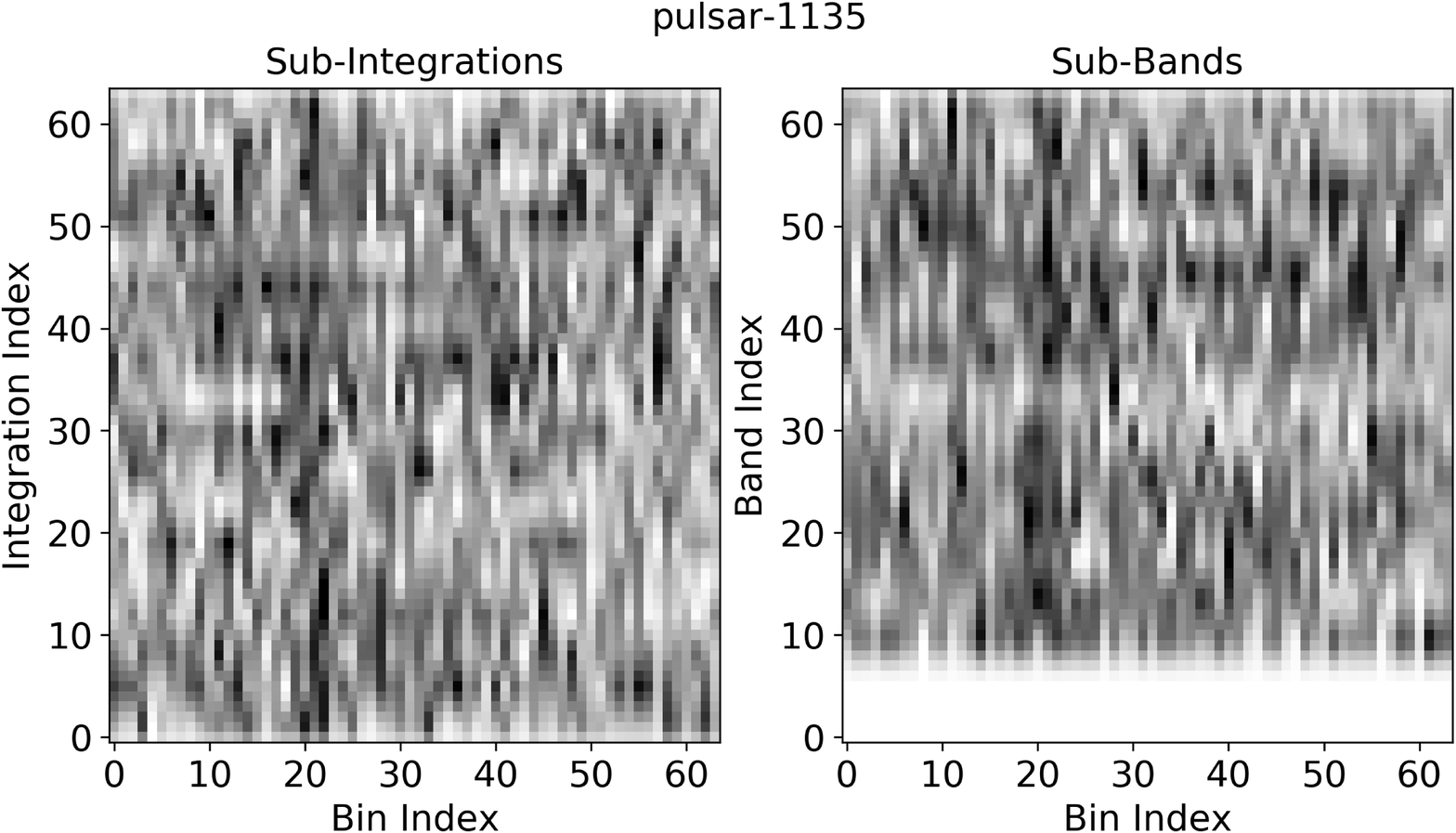}
   \label{p}
  \end{minipage}%

  \begin{minipage}[t]{0.8\textwidth}
  \centering
   \includegraphics[width=0.8\textwidth]{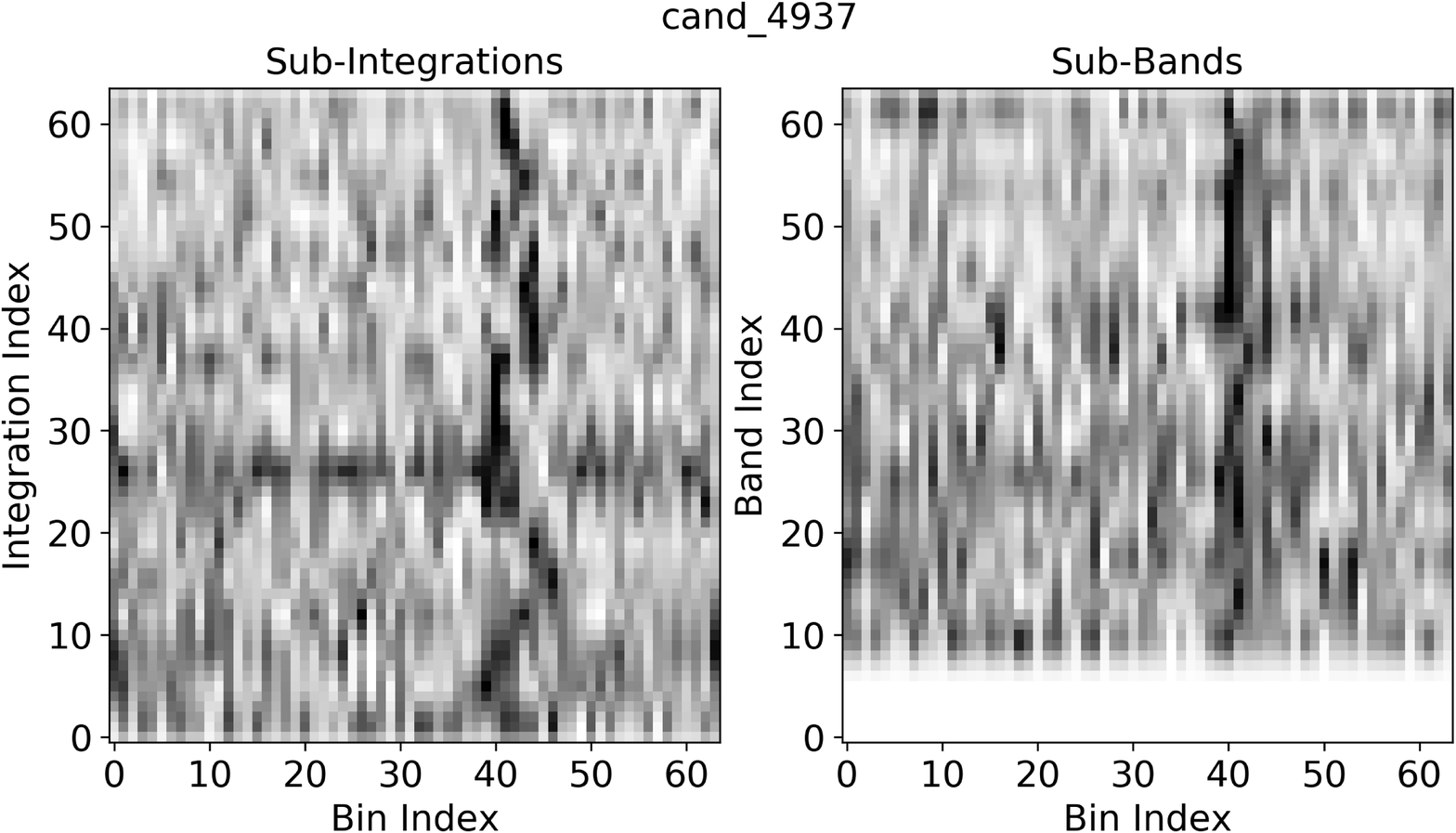}
  \label{c}
  \end{minipage}%

  \caption{{Two mislabeled examples. Subplots in the first row are from  pulsar\_1135, and the orthers are from cand\_4937. Their features in these subplots are not obvious.}}
  \label{Figure10}
\end{figure}

From the testing results, our model has shown the ability to classify pulsar candidates.
We are more concerned with the generalization performance of the model and the classification performance of specific signal types, such as weaker pulsar signals.
We have analyzed the mislabeled testing candidates and found that there are some candidates which are difficult to classify by input subplots.
Figure ~\ref{Figure10} provides two examples of them, one is from pulsar\_1135 while the other is from cand\_4937. With some interference, cand\_4937 is more pulsar-like than the signal of pulsar\_1135.
For these candidates, characteristics of these subplots are not obvious to identify. So if we want to make progress on accuracy, the model needs more resolved information.

\section{Conclusion}

In this work, we have proposed a deep CNN architecture for classification of pulsar candidates.
By using plots as inputs, it is an end-to-end  model and avoids potential bias of artificially designed features.
To address the problem of imbalance, a simple and effective way was designed to synthesize more minority samples, which keep pulsar characteristics.
With the aid of synthetic samples, the model has demonstrated the ability to classifying pulsar candidates in the HTRU 1 dataset.
It can extract features well and classify pulsar candidates reliably. The deep CNN has good application prospects in this field.
To improve generalization performance further, we need to add more abundant and independent data or other valuable information (such as the DM-SNr curve).

\bibliography{ms2019-0001.R1-bibtex}

\begin{thebibliography}{39}
\providecommand\natexlab[1]{#1}
\providecommand\JournalTitle[1]{#1}

\bibitem[Alom {et~al.}(2018)]{Alom2018}
Alom, M.~Z., Taha, T.~M., Yakopcic, C., {et~al.} 2018, arXiv: 1803.01164v1

\bibitem[Bates {et~al.}(2012)]{Bates2012}
Bates, S.~D., Bailes, M., Barsdell, B.~R., {et~al.} 2012, MNRAS, 427, 1052

\bibitem[Bethapudi \& Desai(2018)]{Bethapudi2018}
Bethapudi, S., \& Desai, S. 2018, Astronomy \& Computing, 23, 15

\bibitem[Buda {et~al.}(2018)]{Buda2018}
Buda, M., Maki, A., \& Mazurowski, M.~A. 2018, Neural Networks, 106, 249

\bibitem[Cordes {et~al.}(2004)]{Cordes2004}
Cordes, J.~M., Kramer, M., Lazio, T. J.~W., {et~al.} 2004, New Astronomy
  Reviews, 48, 1413

\bibitem[Der~Maaten \& Hinton(2008)]{Maaten2008}
Der~Maaten, L.~V., \& Hinton, G.~E. 2008, Journal of Machine Learning Research,
  9, 2579

\bibitem[Devine {et~al.}(2016)]{Devine2016}
Devine, T.~R., Gosevapopstojanova, K., \& Mclaughlin, M.~A. 2016, MNRAS, 459,
  1519

\bibitem[Eatough {et~al.}(2010)]{Eatough2010}
Eatough, R.~P., Molkenthin, N., Kramer, M., {et~al.} 2010, MNRAS, 407, 2443

\bibitem[Faulkner {et~al.}(2004)]{Faulkner2004}
Faulkner, A.~J., Stairs, I.~H., Kramer, M., {et~al.} 2004, MNRAS, 355, 147

\bibitem[Ford(2017)]{Ford2017}
Ford, J.~M. 2017, Pulsar Search Using Supervised Machine Learning, PhD thesis,
  Nova Southeastern University

\bibitem[Goodfellow {et~al.}(2016)]{Goodfellow2016}
Goodfellow, I., Bengio, Y., \& Courville, A. 2016, Deep Learning (MIT Press)

\bibitem[Guo {et~al.}(2017)]{Guo2017}
Guo, P., Duan, F., Wang, P., Yao, Y., \& Xin, X. 2017, arXiv preprint
  arXiv:1711.10339V1

\bibitem[He \& Garcia(2009)]{He2009}
He, H., \& Garcia, E.~A. 2009, IEEE Transactions on Knowledge and Data
  Engineering, 21, 1263

\bibitem[He {et~al.}(2016)]{He2016}
He, K., Zhang, X., Ren, S., \& Sun, J. 2016, in Proceedings of the IEEE
  conference on Computer Vision and Pattern Recognition, 770

\bibitem[Hinton {et~al.}(2012)]{Hinton2012}
Hinton, G.~E., Srivastava, N., Krizhevsky, A., Sutskever, I., \& Salakhutdinov,
  R.~R. 2012, Computer Science, 3, 212

\bibitem[Hobbs {et~al.}(2009)]{Hobbs2009}
Hobbs, G.~B., Bailes, M., Bhat, N. D.~R., {et~al.} 2009, Publications of the
  Astronomical Society of Australia, 26, 103

\bibitem[Ioffe \& Szegedy(2015)]{Ioffe2015}
Ioffe, S., \& Szegedy, C. 2015, in International Conference on Machine
  Learning, 448

\bibitem[Johnston {et~al.}(1992)]{Johnston1992}
Johnston, S., Lyne, A.~G., Manchester, R.~N., {et~al.} 1992, MNRAS, 255, 401

\bibitem[Keith {et~al.}(2009)]{Keith2009}
Keith, M.~J., Eatough, R.~P., Lyne, A.~G., {et~al.} 2009, MNRAS, 395, 837

\bibitem[Keith {et~al.}(2010)]{Keith2010}
Keith, M.~J., Jameson, A., Straten, W.~V., {et~al.} 2010, MNRAS, 409, 619

\bibitem[Kingma \& Ba(2015)]{Kingma2015}
Kingma, D.~P., \& Ba, J. 2015, in International Conference on Learning
  Representations, 1

\bibitem[Krizhevsky {et~al.}(2012)]{Krizhevsky2012}
Krizhevsky, A., Sutskever, I., \& Hinton, G.~E. 2012, in Advances in Neural
  Information Processing Systems, 1097

\bibitem[Lee {et~al.}(2013)]{Lee2013}
Lee, K., Stovall, K., Jenet, F., {et~al.} 2013, MNRAS, 433, 688

\bibitem[Lorimer {et~al.}(1998)]{Lorimer1998}
Lorimer, D.~R., Lyne, A.~G., \& Camilo, F. 1998, A\&A, 331, 1002

\bibitem[Lyne {et~al.}(2004)]{Lyne2004}
Lyne, A.~G., Burgay, M., Kramer, M., {et~al.} 2004, Science, 303, 1153

\bibitem[Lyon {et~al.}(2016)]{Lyon2016}
Lyon, R.~J., Stappers, B.~W., Cooper, S., Brooke, J.~M., \& Knowles, J.~D.
  2016, MNRAS, 459, 1104

\bibitem[Manchester {et~al.}(2001)]{Manchester2001}
Manchester, R.~N., Lyne, A.~G., Camilo, F., {et~al.} 2001, MNRAS, 328, 17

\bibitem[Morello {et~al.}(2014)]{Morello2014}
Morello, V., Barr, E.~D., Bailes, M., {et~al.} 2014, MNRAS, 443, 1651

\bibitem[Nan {et~al.}(2011)]{Nan2011}
Nan, R., Li, D., Jin, C., {et~al.} 2011, International Journal of Modern
  Physics D, 20, 989

\bibitem[Nielsen(2015)]{Michael2015}
Nielsen, M.~A. 2015, Neural Networks and Deep Learning (Determination Press)

\bibitem[Radford {et~al.}(2015)]{Radford2015}
Radford, A., Metz, L., \& Chintala, S. 2015, arXiv preprint arXiv:1511.06434

\bibitem[Sabour {et~al.}(2017)]{Sabour2017}
Sabour, S., Frosst, N., \& Hinton, G.~E. 2017, in Advances in Neural
  Information Processing Systems, 3856

\bibitem[Sheikh {et~al.}(2006)]{Sheikh2006}
Sheikh, S.~I., Pines, D.~J., Ray, P.~S., {et~al.} 2006, Journal of Guidance
  Control \& Dynamics, 29, 49

\bibitem[Simonyan \& Zisserman(2015)]{Simonyan2015}
Simonyan, K., \& Zisserman, A. 2015, in International Conference on Learning
  Representations

\bibitem[Smits {et~al.}(2009)]{Smits2009}
Smits, R., Kramer, M., Stappers, B.~W., {et~al.} 2009, A\&A, 493, 1161

\bibitem[Srivastava {et~al.}(2014)]{Srivastava2014}
Srivastava, N., Hinton, G., Krizhevsky, A., Sutskever, I., \& Salakhutdinov, R.
  2014, Journal of Machine Learning Research, 15, 1929

\bibitem[Stokes {et~al.}(1986)]{Stokes1986}
Stokes, G.~H., Segelstein, D.~J., Taylor, J.~H., \& Dewey, R.~J. 1986, ApJ,
  311, 694

\bibitem[Tan {et~al.}(2018)]{Tan2018}
Tan, C.~M., Lyon, R.~J., Stappers, B.~W., Cooper, S., \& Sanidas, S. 2018,
  MNRAS, 474, 4571

\bibitem[Zhu {et~al.}(2014)]{Zhu2014}
Zhu, W., Berndsen, A., Madsen, E., {et~al.} 2014, ApJ, 781, 117

\end{thebibliography}
\bibliographystyle{raa}{}

\end{document}